# MBE obtained n-CdO:Eu/p-Si heterojunctions - electron beam induced profiling, electrical and structural properties

Ewa Przeździecka[1,*], Igor Perlikowski[2], Dawid Jarosz[3], Sergij Chusnutdinow[1], Aleksandra Wierzbicka[1], Abinash Adhikarii[1, 4], Marcin Stachowicz[1], Rafał Jakieła[1], Eunika Zielony[2], Piotr Wojnar[1, 5], A. Kozanecki[1]

* Correspondence: eilczuk@ifpan.edu.pl (E.P.)

**Highlights**
1. XRD patterns reveal that the films have cubic structure with polycrystalline nature.
2. Electron beam induced currents reveal of p-n junction on the CdO and Si interface.
3. The diffusion length of minority carriers were obtained.

**Abstract**

The present investigation reports on the fabrication and characterization of heterojunctions based on in-situ Eu-doped CdO layers, which were deposited on p-type silicon using the plasma-assisted molecular beam epitaxy (PA-MBE) method. The structural and optical properties of the cadmium oxide (CdO) films were investigated using X-ray diffraction and Fourier transform infrared spectroscopy (FTIR). The CdO:Eu films are polycrystalline. The electrical properties of the *p–n* heterojunction composed of transparent *n*-CdO:Eu and *p*-Si semiconductors were investigated by current–voltage and electron beam-induced current (EBIC) measurements. Current-voltage measurements demonstrate good junction characteristics with a rectifying ratio of ~20 at ± 3 V. EBIC measurements allowed us to calculate the diffusion length of minority carriers and the precise location of the depleted area at the CdO and Si interfaces.

## 1. Introduction

Semiconductors based on metal oxides have been studied widely for electronic devices such as solar cells, gas sensors, liquid crystal display, smart windows, flat panel display, optical heaters and light emitting diodes[1,2]. Cadmium oxide (CdO), which is metal-oxide-semiconductor is one of the promising materials for realizing optoelectronic devices [3]. Cadmium oxide is an n-type semiconductor due to oxygen vacancies and has a rock-salt crystalline structure[4–7]. CdO which possesses a direct band gap of 2.3 eV will be a useful material in the optoelectronic applications by making heterojunctions with Si[8,9] GaN [8,9] and CdS[10] and with others *p*-type materials. Many of growth techniques such as sol–gel[11,12], RF sputtering[4,13,14] pulsed laser deposition[15–17], chemical bath deposition[18,19] and molecular beam epitaxy[7,20] have been used to prepare CdO thin films.

Cadmium oxide thin films have usually *n*-type character and interestingly it is possible to tune the electron concentration and mobility in the orders of magnitude by playing with the growth parameters and the stoichiometry of the layers in particular, by manipulating the oxygen deficiency. Understanding of the electrical parameters of CdO layers is crucial for their future applications[4,5,7,21,22] In the case of heterostructures for diodes, it is also important to understand the transport mechanisms at the interface between the *p*-type material and the *n*-type material. The quality of the interface is critical to the transport mechanism and the quality of the diodes[23]. Doping of the layers also affects their physical properties, including electrical properties as was previously observe in the doped CdO layers. [3,15,24–31] To clarify the carrier transport mechanism in *n*-CdO:Eu/*p*-Si junction obtained by MBE, we have investigated the electron-beam-induced current. In particular Eu impact on the minority carrier diffusion length was investigated in this work and the location of junction on the interface was confirmed.

## 2. Experimental

### 2.1 Materials

A series of heterojunctions based on *n*-type Eu-doped CdO layers grown on commercially available *p*-type (100) Si substrates was obtained using the Riber Compact 21B plasma-assisted molecular beam epitaxy (PA-MBE) system. High-purity Cadmium (6N) and Europium (4N) served as sources for Cd and Eu, respectively, within the effusion cells, whereas an oxygen plasma source powered by radio frequency (RF) power was used as a source of oxygen in the MBE setup. Preceding the growth process, the Si (001 oriented) substrates were chemically etched using Buffered Oxide Etch (BOE) for 2 minutes followed by subsequent wet and dry cleaning using deionized water and $N_2$ gas respectively. Then the Si substrates were subjected to annealing within the load chamber of the MBE system at a temperature of 150°C for 1 hour, prior to their transfer to the growth chamber. During the growth, the Cd flux was maintained at approximately $2.2 \times 10^{-7}$ Torr controlled by fixing the Cd effusion cell temperature at 380°C. Meanwhile, the Eu flux was varied from $5.6 \times 10^{-9}$ Torr to $5.0 \times 10^{-9}$ Torr by changing the temperature of the Eu effusion cell (ranging from 300°C to 380°C in increments of 20°C). The growth process was carried out at a temperature of 360°C measured by thermocouple located close to the sample. The oxygen flow was at the level of 3 sccm at a fixed 400W RF power of the oxygen plasma throughout the growth process.

### 2.2 Experimental techniques

The structural analysis of CdO layers doped *in situ* with Eu was carried out using X-ray diffraction (XRD). A Panalytical X'Pert Pro-MRD diffractometer was utilized, featuring a hybrid two-bounce Ge (220) monochromator coupled with an X-ray mirror and a threefold Ge (220) analyzer positioned in frpont of a Pixel detector or proportional detector. All measurements were performed using $Cu_{K\alpha 1}$ radiation with a wavelength of 1.5406 Å.[32,33]

The current voltage (I-V) characteristics are the most important measurements to study the electrical behavior of any diode. I-V measurements were performed using a computer controlled Keithley 4200-SCS semiconductor characterization system in dark. Minority carrier diffusion length measurements were conducted on the structures which were cleaved *in situ* perpendicular to the growth plane in the ZEISS EVO HD15 Scanning Electron Microscope (SEM), using the Digital Image Scanning System (DISS 5) Electron-beam-induced current (EBIC) in the microscope chamber under a 10 kV electron beam accelerating voltage.

FTIR measurements were carried out with infrared spectrometer FTIR Vertex 70v from Brucker equipped with Platinum ATR (Attenuated Total Reflectance) add-on. The samples were placed on the crystal surface and kept in good optical contact, the measurements were performed in vacuum. The background spectra of a blank diamond were used to generate absorption spectra. The acquisition of signal was performed with LN-MCT (HgCdTe) detector cooled with nitrogen. The scan frequency was set to 10 kHz in range of 400-4000 cm$^{-1}$ of the mid-infrared lamp as a emission source. Additionally, a series of experimentation has been performed with Brucker infrared microscope FTIR Lumos II on three selected surface spots of each sample. The 8x objective provides magnification, while the two measurement modes (reflection and ATR) enable the study of both bulk and surface properties of the sample. The ZnSe beam splitter is used to separate the incident and reflected beams in the microscope.

The ATR technique utilizes the phenomenon of total internal reflection, which allows for non-destructive analysis of samples without the need for sample preparation. The diamond crystal in the Platinum ATR accessory enhances the intensity of the reflected light, improving the sensitivity of the measurements. The LN-MCT detector, cooled with liquid nitrogen, provides high-sensitivity and low-noise detection in the mid-infrared range.

## 3. Results and discussion

### 3.1 Structural analysis

The XRD patterns of Eu-doped CdO layers grown on Si (100) substrates are shown in Fig 1. The 400 diffraction peak at about 69º corresponds to the silicon substrate and is clearly distinguishable. The diffraction peak at (*2θ*) about 33º, 38.3º, 55.2º, 65.9º, 81.9º, 91.3º and 94.4º attributed respectively to the 111, 200, 220, 311, 400, 331, and 420 diffraction peaks of CdO cubic rocksalt structure (JCPDF card no. 00-005-0640). All the grown layers are polycrystalline and part of them with the preferred growth direction being expressed in terms of texture coefficient (TC) defined as[34],

$$TC(h,k,l) = \frac{\frac{I(h,k,l)}{I_o(h,k,l)}}{\frac{1}{n}\sum \frac{I(h,k,l)}{I_o(h,k,l)}}$$

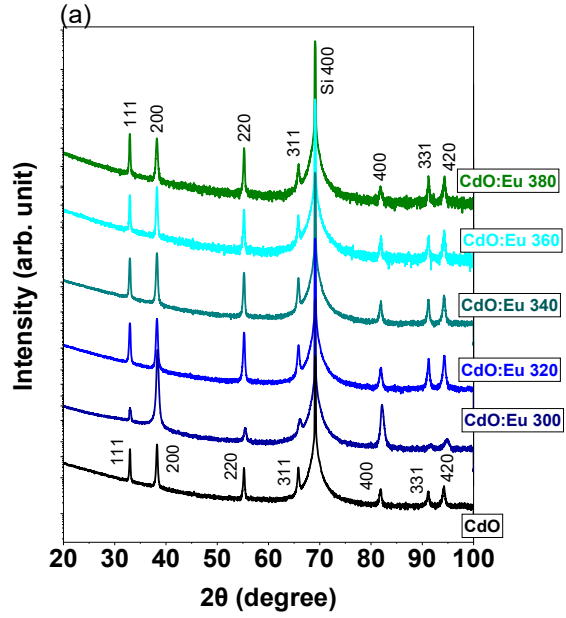

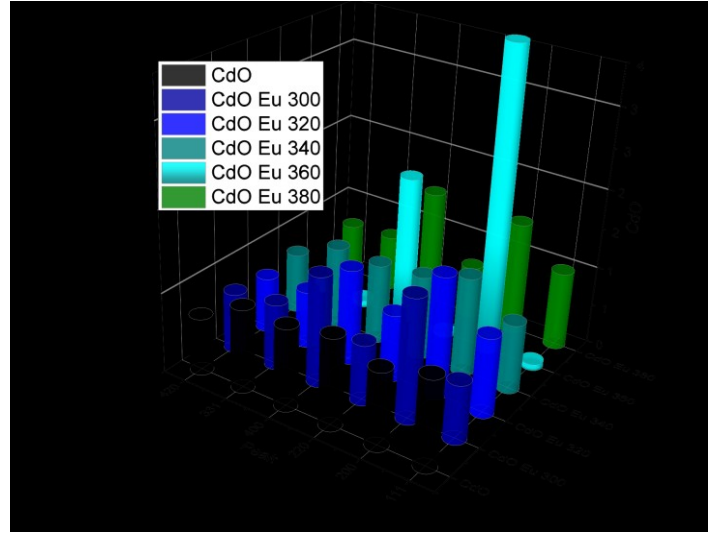

Figure 1. (a) XRD patterns of CdO/*p*-Si films in situ doped with Eu grown by MBE. (b) texture coefficient (TC) analysis for all the X-ray CdO peaks observed in part (a).

where $I$ and $I_o$ are the intensity of diffraction peak and standard intensity based on (JCPDF card no. 00-005-0640) of the *hkl* plane, respectively. The total number of diffraction peaks observed, denoted as $n$, is 6 in the present study. Consequently, the maximum possible value of TC is 6. The calculated value of TC corresponding to all peaks are presented on Fig 1 (b). According to the definition, a TC value close to 1 suggests a randomly oriented layer, while a TC value close to 6 indicates a perfectly oriented layer. TC values greater than 1 suggest a predominance of grains aligned in a specific *hkl* direction. Conversely, values between 0 and 1 indicate a scarcity of grains oriented in that direction [35,36]. The TC values of pure CdO are close to 1 and it increases for 200 peak with an increase in Eu dopant concentration in CdO. In the present study, the highest TC value of 3.788 is obtained for 002 peak for sample CdO:Eu360. This increase in TC value with increasing Eu dopant concentration indicates that the grains are preferentially aligned along the [100] direction. The grain size corresponding to the 200 diffraction peak is determined using Scherrer's relation [37] and listed in Table 1. The grain size of pure CdO is found to be 27.83 nm. The grain size increases with an increase in Eu doping concentration in the CdO lattice and found to be in the range of 29-39 nm.

Table 1. Bragg angle (2θ), FWHM, grain size, and texture coefficient TC of 200 diffraction peak of CdO and Eu-doped CdO layers

|  | $(2\theta)_{200}$ (in degree) | FWHM (in degree) | $D_{200}$ (in nm) | $TC_{200}$ |
|---|---|---|---|---|
| **CdO** | 38.21 | 0.30 | 27.83 | 0.965 |
| **CdO:Eu300** | 38.23 | 0.23 | 36.88 | 1.544 |
| **CdO:Eu320** | 38.23 | 0.25 | 33.35 | 1.527 |
| **CdO:Eu340** | 38.23 | 0.27 | 30.89 | 1.248 |
| **CdO:Eu360** | 38.35 | 0.26 | 32.34 | 3.788 |
| **CdO:Eu380** | 38.25 | 0.22 | 38.38 | 1.442 |

## 3.2 Fourier transform infrared spectroscopy

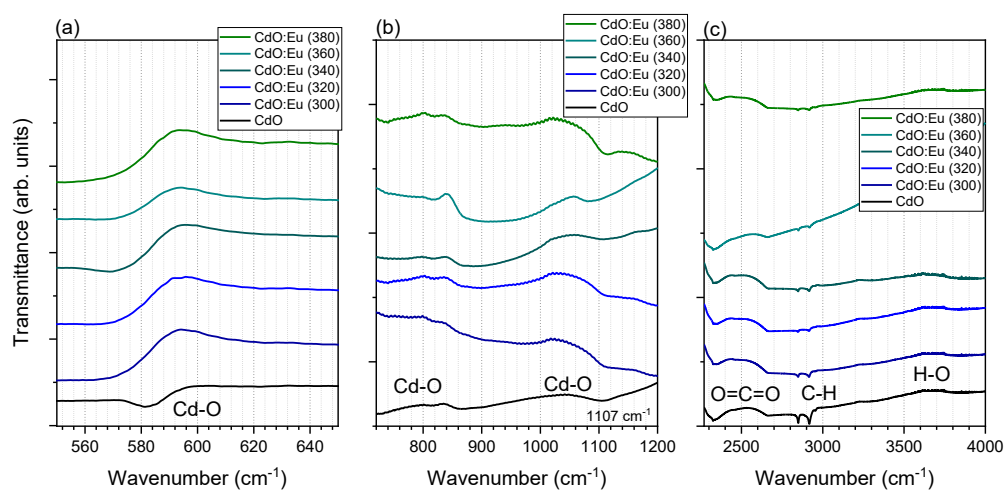

Figure 2. FTIR Transmittance spectra at room temperature.

To study characteristic functional groups observed in the samples, Fourier transform infrared spectroscopy (FTIR) spectra were recorded in the range of 500-4000 cm$^{-1}$ as presented in Fig. 2. For CdO samples the absorption band around 581 cm$^{-1}$ and 839 cm$^{-1}$ and band at about 1117 cm-1 [38] (Fig 2 (a) and 2 (b)) is corresponds to (Cd-O) and confirm the formation of CdO thin film[39]. These line is observed in all the samples, but the shape of the spectrum minority changes with Eu concentration. The broad FTIR absorption band at 3500 -3700 cm-1 is assigned to stretching vibrational mode of a hydroxyl group O– H [39,40]. [38,39] it was also suggested that band band observed at 3612 cm-1 is ascribed to the symmetric stretching mode of H–O–H [41]. The bands at 2930 cm-1 are assigned to stretching modes of C-H group [42] In the literature, the absorption bands at about ~2300 cm-1 are identified as a consequence of the presence of atmospheric carbon dioxide O=C=O. FTIR bands differences modes can be due to development of lattice distortion and/or defects, associated with Eu dopant concentrations.

### 3.3 Electrical characterization

The diodes structures consist of 350–470 nm thick CdO or CdO:Eu layers grown by PA-MBE on commercially available *p*-type Si substrates. The europium doping was confirmed by measured secondary ion mass spectroscopy (SIMS) depth profiles. CdO itself has high carrier (electron) concentration[7], and doping with Eu further increases it[25]. Thus, expected carrier concentration on the *n*-side of the investigated junctions is at the level $1e^{19}$-$1e^{20}$ cm$^{-3}$ and it is ~$2·10^{17}$ cm$^{-3}$ in the case of *p*-Si substrate. In analyzed in this paper junction rectyfying factor at ±3 V is 18 for both CdO/*p*-Si and CdO:Eu/*p*-Si.

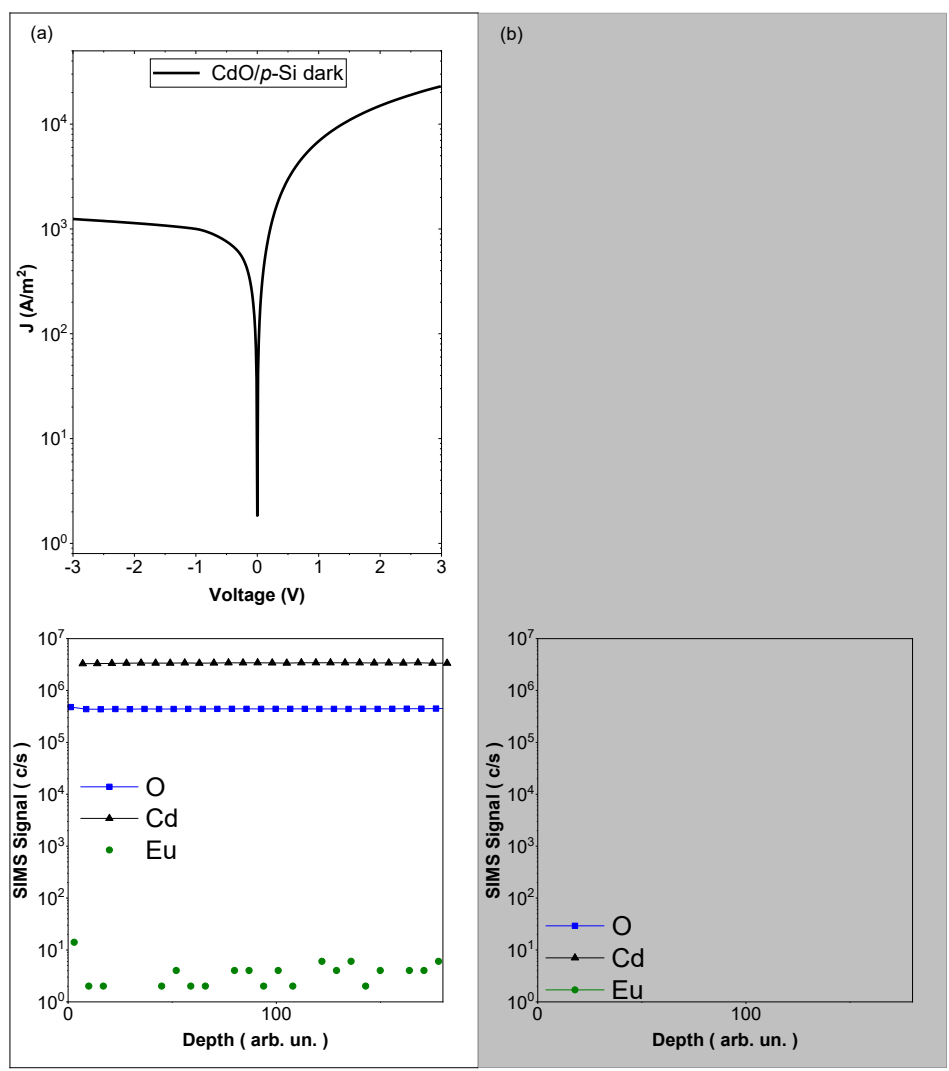

Fig 3. (a) Current density-voltage characteristics of the *n*-CdO/*p*-Si measured at 300K in the dark and SIMS depth profile for this sample (b) Current density-voltage characteristics of the *n*-CdO:Eu/*p*-Si samples measured at 300K in dark and SIMS depth profile for this sample with visible Eu doping profile.

The EBIC technique is the method of choice for determining the minority carrier diffusion length. It is based on measuring the current due to nonequilibrium carriers generated by the electron beam of the scanning electron microscope (SEM) and collected by the built-in field of the barrier. The technique can been used to determine carrier lifetime, diffusion length, defect energy levels, and surface recombination velocities. Charge collection images with simultaneously gathered SEM images reveal the location of p-n junctions, recombination sites such as dislocations and precipitates, and the presence of doping level inhomogeneity [43–47]. In Fig. 4, cross sectional SEM image of a typical CdO:Eu/Si structure cleaved perpendicular to the growth plane is presented. The EBIC signal is visible as a yellow color and also EBIC line scan has been superimposed on cross-sectional SEM image. As it is seen, the maximum EBIC signal corresponds to CdO:Eu and Si interface. The EBIC line profile presented in Figure 4 is characteristic of a p- n device. It is evident that the junctions exists on the CdO/Si interface and are continuous. The minority carrier diffusion lengths can be extracted from EBIC line scan according to the following relation:

$$I_{EBIC} = I_0 \exp(-x/L_{e,p})$$
(1)

where $I_{EBIC}$ is the EBIC signal, $I_0$ is a constant, $x$ describes the position of the generating electron beam. The data fitting based on this relation is represented by the yellow curves in Figure 4.

The diffusion lengths, extracted from EBIC line scans are 100 nm and 165 nm for minority electrons for CdO and CdO:Eu layers, respectively.

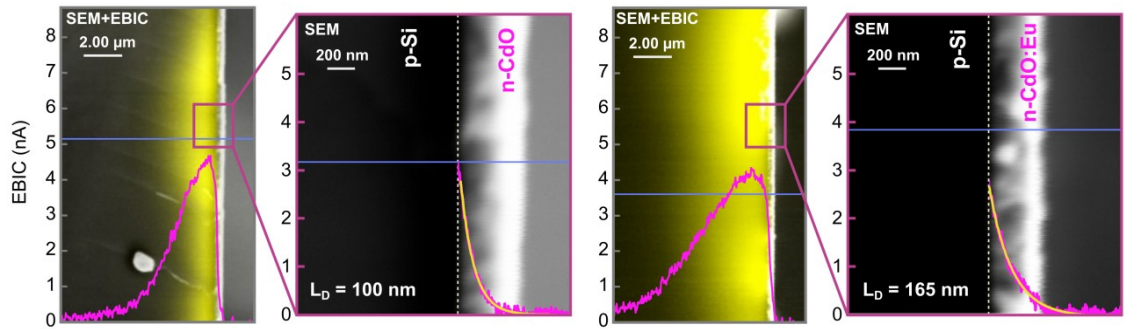

Figure 4. Cross-sectional SEM image of the CdO:Eu/Si interfaces, EBIC line scans are superimposed on the SEM image.

The diffusion length value in the previously reported oxide based structure *n*-MgZnO/*p*-AlGaN/GaN was 120 nm for holes in *n*-MgZnO and 890 nm for electrons in *p*-AlGaN/GaN regions [48]while in the case of the n-ZnO/n-GaN structure minority carrier diffusion length was reported in the range of 125–175 nm [[49]. In case of *n*-CdO/electrolyte junction high value of the hole diffusion length at about $10^{-4}$ cm has been calculated by the Butler-Gartner equation from measurements of photocurrent quantum efficiency[50]. For example, for other structures like CuS/CdS it was 90 to 1700 nm for holes in *n*-type CdS and 110 to 570 for electrons in *p*-type CuS [51] whereas for PbSe/CdTe it was 1000 nm and 1270 nm for minority holes and electrons, respectively [52]. In conventional p–n junctions, a long minority carrier diffusion length is essential; thus, good-quality, well-ordered semiconductors are required to produce high-quality junctions. In our case, we observe a higher carrier diffusion length in CdO layers doped with Eu, which can be correlated with the preferential orientation observed in these CdO:Eu layers.

**Conclusions**

Thin films of cadmium oxide (CdO) doped with Eu were prepared by plasma assisted molecular beam epitaxy on *p*-type 0001 Si substrates. The XRD diffraction patterns confirms the formation of polycrystalline CdO films in the cubic crystal structure. The peaks related to Cd–O vibrations were observed from FT-IR studies. From the electrical studies, it was observed that *p-n* junction with rectification ratio about 20 was prepared. The homogenies Eu doping in CdO layers during the epitaxy was confirmed by SIMS depth profile measurements. By applying SEM and E-BIC measurement the location of the p-n junction on the CdO and Si interfaces was confirmed. The minority carrier diffusion length at about 100 nm and 165 nm in case of CdO and CdO doped with Eu were respectively detected.

**Conflicts of Interest:** Author Ewa Przezdziecka has received research grants from the Polish National Science Center. Author Ewa Przezdziecka, Aleksandra Wierzbicka, Abinash Adhikarii and Igor Perlikowski have received an honorarium from this grant. The funders had no role in the design of the study; in the collection, analyses, or interpretation of data; in the writing of the manuscript; or in the decision to publish the results
The authors declare that they have no known competing financial interests or personal relationships that could have appeared to influence the work reported in this paper.

**CRediT authorship contribution statement**

 **Ewa Przezdziecka:** Conceptualization, Funding acquisition, Supervision, Project administration, Writing – original draft. **Igor Perlikowski:** Formal analysis, Investigation, Visualization, Writing – review and editing. **Dawid Jarosz:** Investigation. **Sergij Chusnutdinow:** Methodology, Investigation. **Aleksandra Wierzbicka:** Methodology, Investigation. **Abinash Adhikari:** Data curation, Formal analysis, Writing – original draft. **Marcin Stachowicz:** Formal analysis, Writing – original draft. **Rafał Jakieła:** Investigation. **Eunika Zielony:** Writing – review and editing, Supervision. **Piotr Wojnar:** Writing – review and editing. **Adrian Kozanecki:** Writing – review and editing.

**Acknowledgments**
This work was supported in part by the Polish National Science Center, Grants No. 2021/41/B/ST5/00216

**Data Availability**
The original contributions presented in the study are included in the article; further inquiries can be directed to the author.